# Temporal interfaces by instantaneously varying boundary conditions

Luca Stefanini[1,†], Shixiong Yin[2,3], Davide Ramaccia[1,†], Andrea Alù[2,3],

Alessandro Toscano[1] and Filiberto Bilotti[1]

[1] *"Roma Tre" University, Via Vito Volterra 62, 00146, Rome, Italy*

[2] *Photonics Initiative, Advanced Science Research Center, New York, NY 10031, USA*

[3] *City College, City University of New York, New York, NY 10031, USA*

**Abstract**

Temporal metamaterials have been recently exploited as a novel platform for conceiving several electromagnetic and optical devices based on the anomalous scattering response arising at a single or multiple sudden changes of the material properties. However, they are difficult to implement in realistic scenarios by switching the permittivity of a material in time, and new strategies to achieve time interfaces in a feasible manner must be identified. In this paper, we investigate the possibility to realize a temporal metamaterial without acting on the material properties, but rather on the effective refractive index and wave impedance perceived by the wave during the propagation in an empty guiding structure by varying the boundaries in time. We demonstrate analytically and through numerical experiments that suddenly changing the physical distance between the metallic plates of a parallel-plate waveguide will induce an effective temporal interface. In addition to the standard backward and forward scattered fields at different frequencies due to the temporal interface, we also identify the presence of a static field necessary to satisfy the continuity of the electromagnetic field across the interface. The proposed concept can be extended to temporally controlled metasurfaces, opening an easier path to the design and realization of novel devices based on time-varying metamaterials at microwave and optical frequencies.

---

a) †To whom correspondence should be addressed: †luca.stefanini@uniroma3.it, davide.ramaccia@uniroma3.it.

# 1  INTRODUCTION

Temporal metamaterials are artificial electromagnetic materials whose effective properties vary over time. In case of abrupt changes of the material properties, the propagating wave experiences a scattering process somewhat analogous to the one happening at the spatial interface between two different media, realizing the temporal counterpart of a spatial discontinuity. This configuration is referred to as a *temporal interface* and is characterized by the generation of a backward and a forward wave, whose amplitudes and frequencies are related to the jump of the refractive index and wave impedance between the two media, *i.e.,* before and after the changing of medium properties [1], [2]. The possibility to control the scattering by also using the time dimension has drawn significant attention in recent years [3]–[6]. Dual to spatially engineered materials, tailoring the material electromagnetic properties in time enables plenty of intriguing wave phenomena [7], *e.g.,* inverse prism [8], temporal aiming [9], synthesis of effective media [10], [11], temporal parity-time symmetry [12], and temporal Brewster angle [13], as well as novel devices in time-domain, *e.g.* antireflection coatings [14], [15], temporal Fabry-Perot cavities [16], photonic time crystals [17], and broadband absorbers [18]–[20]. Proposal on how to realize a temporal interface for electromagnetic waves have been issued in transmission line structures [19], and experimentally demonstrated in plasmas [21] and graphene [22], where the bulk material properties were suddenly modified. However, all the proposed techniques so far are difficult to implement in realistic scenarios, requiring changing instantaneously the properties of the whole medium where the propagation is taking place. Recently, Miyamaru et al. [23] have experimentally observed that a frequency shift is achieved when a laser pulse changes the carrier density of a dielectric layer where an electromagnetic wave is propagating. The change of the surface properties induces a scattering process like the one observed at a temporal interface. The feasibility of temporal metamaterials is currently limited by the low implementation strategies that overcome the requirement to change the properties of the whole medium where the propagation is taking place. Novel approaches for exploiting the anomalous scattering by temporal discontinuities must be identified.

In this work, we investigate the possibility to realize a temporal interface without acting on the actual material properties, but rather on the effective ones. Quantities like the effective refractive index and effective wave impedance are typically used for describing the propagation characteristics within a guiding system, analogous to the wave propagation in an unbounded medium with the same effective material properties. Here, we consider an empty parallel-plate waveguide (PPWG) whose boundaries are suddenly changed in time to induce an effective temporal interface. Such an approach allows relaxing the step-like modulation in time of the bulk properties of the filling material and using the change of the bounding surfaces as degree of freedom. Moreover, parallel-plate guiding systems have been already demonstrated to be an effective strategy to achieve artificial materials with negative and near-zero permittivity [24], which realizes a good platform to investigate the wave dynamics of time-switching in metamaterials. In Fig. 1, we illustrate two possible strategies for enabling boundary-induced interfaces within a guiding structure: *i)* changing the properties of the bounding surfaces of the waveguide, represented by the arbitrary surface impedance $Z$ (Fig. 1a-c); or *ii)* changing the distance between two surfaces with $Z=0$, *i.e.*, ideal metallic plates (Fig. 1b-d). The former represents an interesting strategy for actual implementation of the boundary-induced interfaces, exploiting the recent progress in engineering the properties of metasurfaces. For example, Fig. 1a shows the case of two waveguides bounded by different metasurfaces realizing a spatial interface, and Fig. 1c shows its temporal counterpart, where the surface properties are instantaneously changed over time, switching from $Z_1$ to $Z_2$ at $t=t_0$ [25]–[29]. This ensures the continuity of the fields across the temporal interface, being only the boundaries modified. The analysis of this configuration can be performed starting from the derivation of the eigenmodes of a metasurface-bounded waveguide for a given pair of metasurfaces, as derived recently by Ma *et al.* in [30]. However, the latter strategy explores the physical process triggered by an abrupt modification of the boundary conditions over time in a more familiar manner, allowing to catch all the interesting features of a temporal interface induced through the variation of the boundary conditions more easily. In particular, the ideal metallic plates in Fig. 1d are assumed to change instantaneously their position at the instant $t=t_0$ from $d_1$ to $d_2$. Although a continuous movement of the metallic plates can be demonstrated, as in [31]–[34], another way to emulate the change in boundary conditions is to assume that the conductivity of the original plates with separation $d_1$ is switched from infinity to 0, while a new pair of metallic plates at are created separated by $d_2$ suddenly. It is expected that the effective refractive index perceived by the wave should change abruptly in time, preserving the momentum and thus maintaining the spatial distribution of fields across the interface [1], [2], [15].

The paper is organized as follows. In Section 2, we first analytically derive the frequency conversion induced by the temporal interface as a function of the jump of the effective refractive index perceived by the wave before and after the boundary induced temporal interface, strongly related to the change of the inter-plate distance; then, by imposing the continuity of the fields across the temporal interface, we derive the scattering coefficients for the forward and backward propagating modes. Moreover, the presence of a static field together with a dynamic one is demonstrated. Section 3 is dedicated to the numerical experiments and to their comparison with the analytical results. Finally, in Section 4, some conclusions are drawn.

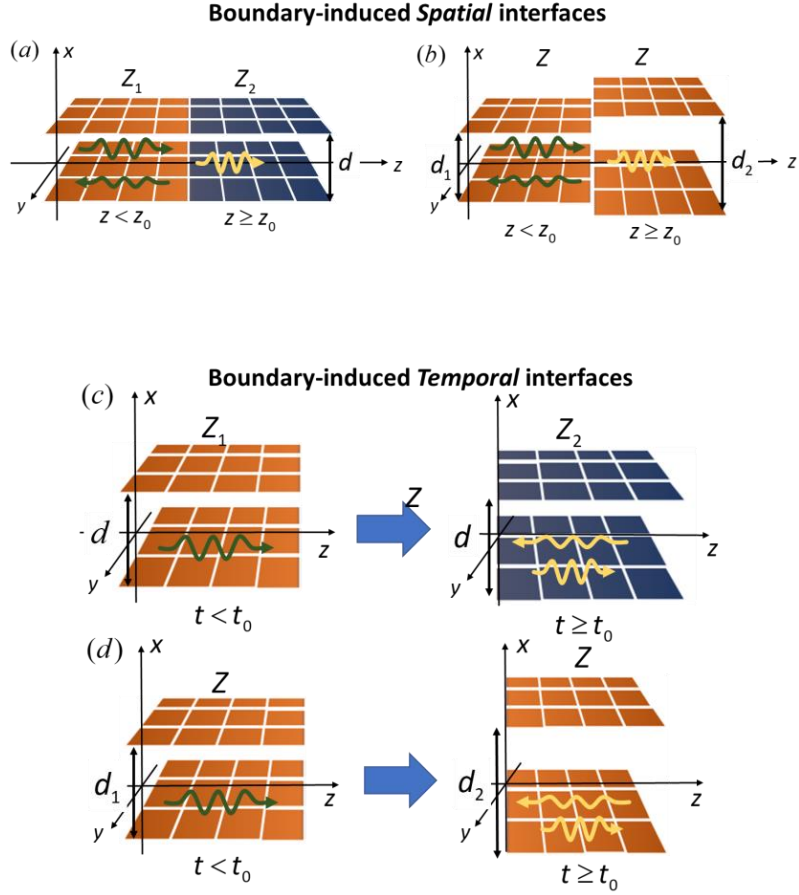

Fig. 1. Graphical illustration of the wave propagation discontinuities induced by the boundaries in a guiding structure: (a) spatial and (c) temporal interfaces induced by a change of the bounding surface impedance; (b) spatial and (d) temporal interfaces induced by a change in the separation between waveguide plates.

## 2 ANALYSIS OF BOUNDARY-INDUCED TEMPORAL INTERFACE

The scattering arising at an interface between two different media is a basic phenomenon in optics and electromagnetics. It takes place not only in presence of an actual change of the medium parameters, *i.e.*, refractive index and wave impedance, but also at the terminals between two different waveguides (see Fig. 1b), as described by the well-established guided wave theory [35]. This allows extending the concept of *interface* to any spatial or temporal locus where the conservation of the wave quantities, *i.e.*, electric, and magnetic fields, is not satisfied anymore. In this framework, two different waveguides support two separate sets of eigenmodes and the matching of the mentioned quantities at the discontinuity location $z_0$ can be achieved only through the generation of a reflected (backward-propagating) and transmitted (forward-propagating) wave. Fig. 1d illustrates the same abrupt change of the waveguide dimensions but in the time domain. Before the switching time $t = t_0$, the propagating mode is defined by the excitation frequency $\omega_1 = 2\pi f_1$, the waveguide dimension $d_1$, and the electromagnetic properties of the filling medium. At the switching time $t = t_0$, the waveguide dimension suddenly changes to $d_2$, keeping the same symmetry, leading to a mismatch between the actual propagating mode and the ones supported by the waveguide in the new configuration. However, the scattering process at a temporal interface is deeply different with respect to the one observed at a spatial interface. In the following, we describe the considered scenario based on a parallel plate waveguide, where the temporal interface induced by the boundaries is realized. The scattering problem at the interface is analytically described, showing that in addition to the expected forward and backward scattered fields, an electrostatic field is excited such that the continuity of the total electromagnetic field is conserved across the interface.

## *2.1 Frequency conversion induced by boundary-induced temporal interface*

Let us consider an infinitely extended parallel-plate waveguide (PPWG) consisting of two metallic plates separated by a distance *d* in the *x*-direction and supporting the propagation of a *z*-directed wave, as shown in Fig. 2a. The waveguide is filled with vacuum (*n*=1). The propagation of the fundamental TEM mode is always supported at any frequency, whereas the TE/TM modes are supported beyond the corresponding cut-off frequencies, dictated by the electrical dimension $d/\lambda$. In this scenario, since the TEM mode always travels with the same phase velocity along the waveguide, regardless the distance *d*, it is insensitive to the effective change of refractive index induced by the temporal discontinuity. We focus therefore on excitations of TE/TM modes. In addition, provided that our boundary modification preserves certain symmetries, TE/TM modes remain orthogonal to each other, allowing us to consider only one of the two sets, *e.g.*, TM modes. Similar discussions and results can be derived for the TE mode case by duality and are not presented in this work.

As shown in Fig. 2a, the TM$_1$ mode is propagating within a PPWG. The propagation vector $\mathbf{k}_z$ is the projection along the z-axis of the filling medium wavevector $\mathbf{k}_0$, whose magnitude is $k_z = \sqrt{k_0^2 - k_x^2}$, where $k_0 = \omega/c = 2\pi/\lambda$ is the magnitude of wavenumber, and $k_x = \pi/d$ is the transverse wavenumber imposed by the waveguide dimensions.

In this scenario, the effective refractive index perceived by the guided mode is [36]:

$$n_{eff} = \sqrt{1-\left(k_x/k_0\right)^2} = \sqrt{1-\left(\lambda/2d\right)^2} \, , \tag{1}$$

and its phase velocity is $v_p = \omega/k_z = \omega/\sqrt{k_0^2 - \left(\pi/d\right)^2}$.

Here, we analyse the frequency conversion expected by an abrupt change of the waveguide dimension, which, in turn, modifies the effective medium perceived by the propagating mode. Fig. 2b reports the dispersion diagram at the temporal interface (Fig.1d), where we can identify the excited modes in the parallel plate waveguide. Before the switching time $t_0$, the fundamental TM$_1$ mode propagates with a propagation wavenumber $k_{z0}$ and a transverse wavenumber $k_{x0}$, represented by the dashed black lines in Fig. 2b.

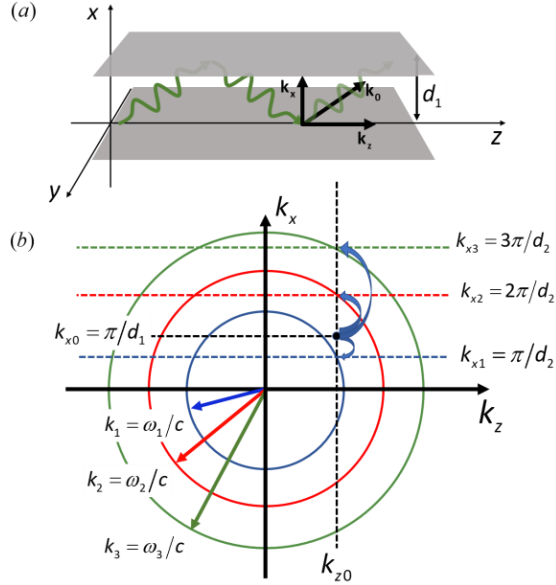

Fig. 2. (a) Representation of the propagation quantities of the TE/TM modes within a parallel-plate waveguide of dimension $d$. (b) Dispersion diagram of the excited modes supported by the waveguide after a temporal discontinuity induced by a sudden change of the waveguide dimensions from $d = d_1$ to $d = d_2$.

At the temporal discontinuity, $t = t_0$, the original propagating mode TM$_1$ perceives an instantaneous change of the effective refractive index from $n_{eff}(d_1)$ to $n_{eff}(d_2)$, which induces the instantaneous change of the temporal frequency through the scaling time-dilation factor $\xi$ [7], [15]:

$$\omega_1 = \xi\omega_0, \quad \text{where} \quad \xi = \frac{n_{eff}(d_1)}{n_{eff}(d_2)} = \sqrt{\frac{1-(k_{x0}/k_0)^2}{1-(k_{x1}/k_1)^2}}, \quad (2)$$

and the generation of a backward and a forward scattered wave, whose relative amplitudes with respect to the original mode will be derived in the next Sections (Sec. 2c-d).

Here, we focus our attention on the temporal and spatial wave quantities across the interface represented on the dispersion diagram in Fig. 2b. A temporal interface preserves the propagating wavelength λ [1], which is directly related to the propagating wavevector $k_z$. In this scenario, in the same manner as a frequency source is represented by a circle in the dispersion diagram, the existing field acts as a wavelength source and is represented by the vertical black dashed line in Fig. 2b. Therefore, in the new waveguide we have excited not only the first TM$_1$ mode, but also all the TM modes defined by the transverse wavenumbers $k_{xm} = m\pi/d_2$ with $m = 1, 2, 3, ...$, and the same original propagation wavenumber $k_z = k_{z0}$ [36]. In Fig. 2b, the circles crossing the points of intersection between $k_{z0}$ and $k_{xm}$ represent the wavevectors $k_m$ of the new modes propagating in the waveguide, each of which has a different frequency $\omega_m = ck_m$. By imposing the continuity of the propagating wavenumber $k_{z0} = k_{zm}$, we analytically derive the new frequencies after the boundary-induced temporal interface:

$$\omega_m = \omega_0 \sqrt{1 - \frac{k_{x0}^2 - k_{xm}^2}{k_0^2}}, \quad m = 1, 2, 3, ... \tag{3}$$

## 2.2 Evaluation of the scattered fields

In this section, we focus our attention on the scattering process taking place at the boundary-induced temporal interface within a parallel plate waveguide and the analytical derivation of the amplitudes of the scattered fields in backward and forward direction after the interface. The most important relationship to be conserved at the interface is the continuity on the induction fields **B** and **D** [1]. In our scenario, being the bulk medium always vacuum, the condition on **B**/**D** relaxes on **H**/**E** as follow for the TM mode configuration under analysis:

$$\begin{cases} H_y(x,z)^{t_0^-} = H_y(x,z)^{t_0^+} \\ E_x(x,z)^{t_0^-} = E_x(x,z)^{t_0^+} \\ E_z(x,z)^{t_0^-} = E_z(x,z)^{t_0^+} \end{cases}. \tag{4}$$

In Fig. 3 we report the components $H_y, E_x, E_z$ of the propagating field just before and after the temporal interface. At the instant of time $t = t_0^+$, the mode can maintain its original profile only if a discontinuity is present where the metallic walls laid before the switching time, leading to a non-null local divergence at those loci. At the temporal interface, the sudden disappearance of metallic plates does not annihilate or take out the surface charges and corresponding surface currents at $t = t_0^-$, as we assumed the abrupt changes in conductivity rather than moving the plates which requires external mechanical energy. In this case, we must face an electrostatic potential problem before moving towards the evaluation of the actual scattering parameters of the electrodynamic fields. It is clear that the fields in the right-hand side of **Error! Reference source not found.** are a superposition of electrostatic and electrodynamic (propagating) fields. Let us start evaluating the electrostatic fields configurations raising at the temporal interface. The static fields $E_{x,z}^{(s)}$ can be evaluated from the electrostatic potential $\Phi(x,z)$ generated by the charges as

$$E_{x,z}^{(s)}(x,z) = \partial_{x,z}\Phi(x,z), \tag{5}$$

where the electrostatic potential is defined by integral of the Green's function over the charge distribution as:

$$\Phi(x,z) = \int_{-\infty}^{+\infty} G(x,x',z,z')\rho(x',z')dx'dz', \tag{6}$$

where $x', z'$ are the coordinates of the charge distribution. Using in (6) the Green's function for the 2D Laplacian and a charge distribution $\rho(x',z')$, we can evaluate the spatial profile within the three regions shown in Fig. 3(d)-(f) for the electrostatic field raising at the temporal interface (see Supplemental Material):

$$E_x^{(s)} = -E_{x0}^{(s)} \times \begin{cases} 2\sinh(k_z(x-d_1/2))e^{-k_z(d_1/2)}e^{-ik_z z}, & \text{within Region 0} \\ 2\cosh(k_z(x+(d_2-d_1)/2))e^{-k_z(d_2-d_1)/2}e^{-ik_z z}, & \text{within Region 1}^- \\ -2\cosh(k_z(x-(d_1+d_2)/2))e^{-k_z(d_1+d_2)/2}e^{-ik_z z}, & \text{within Region 1}^+ \end{cases}$$

$$E_z^{(s)} = -E_{z0}^{(s)} \times \begin{cases} 2\cosh(k_z(x-d_1/2))e^{-k_z(d_1/2)}e^{-ik_z z}, & \text{within Region 0} \\ 2\sinh(k_z(x+(d_2-d_1)/2))e^{-k_z(d_2-d_1)/2}e^{-ik_z z}, & \text{within Region 1}^- \\ -2\sinh(k_z(x-(d_1+d_2)/2))e^{-k_z(d_1+d_2)/2}e^{-ik_z z}, & \text{within Region 1}^+ \end{cases} \tag{7}$$

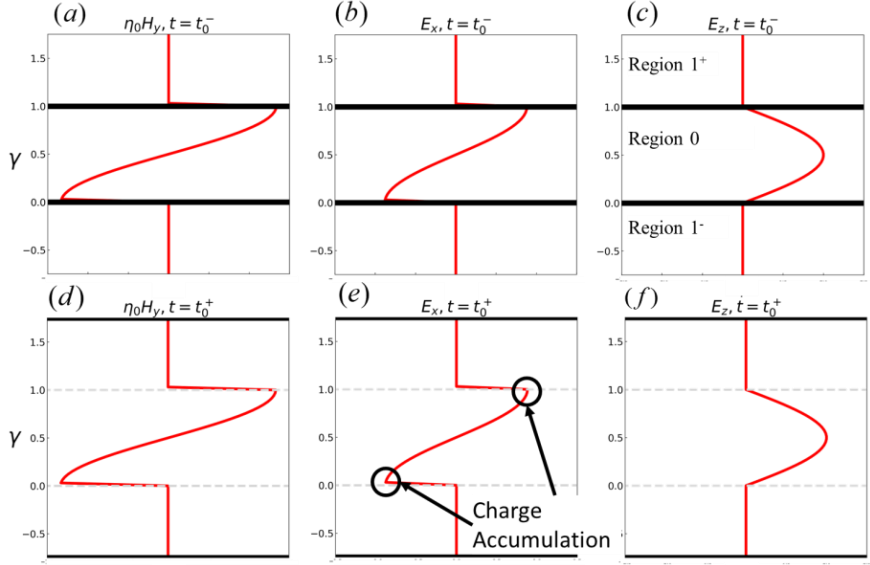

Fig.3. Field configuration for the $H_y, E_x, E_z$ components before (a-c) the temporal interface at $t = t_0^-$ and same configurations (d-f) after the temporal interface at $t = t_0^+$. The domain is divided in Region 0, inside of the initial PPWG, and Region 1$^+$/1$^-$ for the upper and lower space between the initial PPWG and the new one. The metallic plates, bulk black, change their position and the field inside remains unaltered, in light gray (d-f) the previous position is indicated. In (e) the field discontinuity is circled in black, at those points there is a charge accumulation.

We can continue evaluating the electrodynamic scattered fields at the interface. As discussed in Sec. 2.2, after the temporal interface the original mode is coupled with all the modes supported by the wider waveguide with the same propagating wavenumber of the original one. However, now, we know that also an electrostatic field is present and contribute to the total field after the interface as follow:

$$\begin{cases} H_y^{t_0^+}(x,z) = \sum_{m=1}^{+\infty} \left( a_m e^{i\omega_m t} - b_m e^{-i\omega_m t} \right) H_{y,m}^{t_0^+} \\ E_x^{t_0^+}(x,z) = E_x^{(s)} + \sum_{m=1}^{+\infty} \left( a_m e^{i\omega_m t} + b_m e^{-i\omega_m t} \right) E_{x,m}^{t_0^+} \\ E_z^{t_0^+}(x,z) = E_z^{(s)} + \sum_{m=1}^{+\infty} \left( a_m e^{i\omega_m t} + b_m e^{-i\omega_m t} \right) E_{z,m}^{t_0^+} \end{cases} \quad (8)$$

where $a_n, b_n$ are the scattering coefficients of the forward-propagating (at frequency $+\omega_m$), and backward-propagating (at frequency $-\omega_m$) modes.

The propagation of the electromagnetic field across the boundary-induced temporal interface is still governed by the Maxwell equations, that for the TM modes under analysis can be casted as

$$\begin{cases} \partial_z H_y^t = -\epsilon_0 \partial_t E_x^t \\ \partial_x H_y^t = \epsilon_0 \partial_t E_z^t \\ \partial_x E_z^t - \partial_z E_x^t = \mu_0 \partial_t H_y^t \end{cases} \quad (9)$$

In (9), the field configuration before ($t = t_0^-$) and after ($t = t_0^+$) the temporal interface can be arbitrarily substituted in virtue of the continuity conditions of the fields across the interface as in **Error! Reference source not found.**. In particular, by using the field expression in (8) in the right-hand side and the original mode in the left-hand side of (9), we obtain:

$$\begin{cases} \partial_z H_y^{t_0^-} = -\epsilon_0 \partial_t \sum_{m=1}^{+\infty} \left( a_m e^{i\omega_m t} + b_m e^{-i\omega_m t} \right) E_{x,m}^{t_0^+} \\ \partial_x H_y^{t_0^-} = \epsilon_0 \partial_t \sum_{m=1}^{+\infty} \left( a_m e^{i\omega_m t} + b_m e^{-i\omega_m t} \right) E_{z,m}^{t_0^+} \\ \partial_x E_z^{t_0^-} - \partial_z E_x^{t_0^-} = \mu_0 \partial_t \sum_{m=1}^{+\infty} \left( a_m e^{i\omega_m t} - b_m e^{-i\omega_m t} \right) H_{y,m}^{t_0^+} \end{cases} \quad (10)$$

The equation system (10) isolates the electrodynamic problem from the electrostatic one, allowing deriving the forward and backward scattering coefficient in a straightforward manner. In particular, the coefficients can be easily retrieved using the mode matching technique: invoking the orthogonality of the modes on the right-hand side is possible to isolate each element by multiplying for the desired $m^{th}$ configuration and integrate over the waveguide's section (more information can be found in the Supplemental Material).

Finally, the scattering coefficients $FW_m$ and $BW_m$ for $m^{th}$ propagating mode are:

$$\begin{cases} a_m = FW_m = \phantom{-}\dfrac{1}{2}\dfrac{\omega_0}{\omega_m}\dfrac{k_{xm}}{k_{x0}}\left(1+\dfrac{\omega_0}{\omega_m}\right)\kappa_m \\ b_m = BW_m = -\dfrac{1}{2}\dfrac{\omega_0}{\omega_m}\dfrac{k_{xm}}{k_{x0}}\left(1-\dfrac{\omega_0}{\omega_m}\right)\kappa_m \end{cases} \quad (11)$$

where $\kappa_m$ is the overlap integral between the initial $TM_1$ and the final $TM_m$ modes (See Supplemental Material for derivation).

# 3 NUMERICAL EXAMPLES

In this Section, we report numerical simulations carried out by using a Finite-Difference Time-Domain (FDTD) technique able to simulate the propagation of TM modes within a parallel-plate waveguide and evaluate in real-time the electrodynamic fields and the electrostatic field induced at the temporal interface due to the abrupt change of the distance between the metallic plates. In all simulations, the waveguide is excited by a narrowband Gaussian pulse at frequency $\omega_0$ for ensuring the finite energy within the domain during the simulation but maintaining a narrow spectrum for properly computing the scattering coefficients The metallic plates are implemented via Dirichlet condition for the longitudinal component of the electric field, ($E_z = 0$ on the metallic plates, see Fig. 2(a)), and via Neumann conditions for the transverse component of the electric field, ($\partial_x E_x = 0$ on the metallic plates, see Fig. 2(a)), and for the magnetic field, ($\partial_x H_y = 0$ on the metallic plates, see Fig. 2(a)) [37]. At the instant $t = t_0$, the aforementioned boundary conditions are not applied anymore on the original walls' position, *i.e.*, $x = 0, d_1$, but on the new ones, *i.e.*, $x = -\alpha, d_1 + \alpha$, where $\alpha = 0.5(d_2 - d_1)$ (see Supplemental Material). Fig. 4 and Fig. 5 report the main numerical results and comparison with the analytical expression of Sec. 2 for the electrodynamic and electrostatic fields, respectively.

As for the electrodynamic fields, Fig. 4a shows three snapshots in time of the magnetic field distribution within the guiding structure whose dimension switches from $d_1$ to $d_2 = 2d_1$ at the instant of time $t = t_0$. As expected, just after the temporal interface (central frame in Fig. 4(a)), the field configuration does not change satisfying the continuity of the E- and H-field at that interface. For $t > t_0$ (right frame in Fig. 4(a)), the field within the waveguide is a superposition of several TM modes filling the entire waveguide, as demonstrated in Fig. 4(b), where the spectra of the original (orange dashed line), backward-propagating (solid light-blue line) and forward-propagating (solid blue line) are reported. The different excited modes propagating at the new frequencies $\omega_m$ are clearly shown. Fig. 4(c)-(e) compare the analytical and numerical excited frequencies, backward and forward scattering coefficients for the first three modes excited in the new waveguide after the boundary-induced temporal interface as a function of the parameter $\gamma = d_2 / d_1$. These modes receive almost all the energy carried by the original TM$_1$ propagating in the waveguide before the switching time $t_0$ due to their lower phase mismatch between the original propagating and final supported configurations. The agreement between numerical and analytical results is very good, confirming the process illustrated in Fig. 2b.

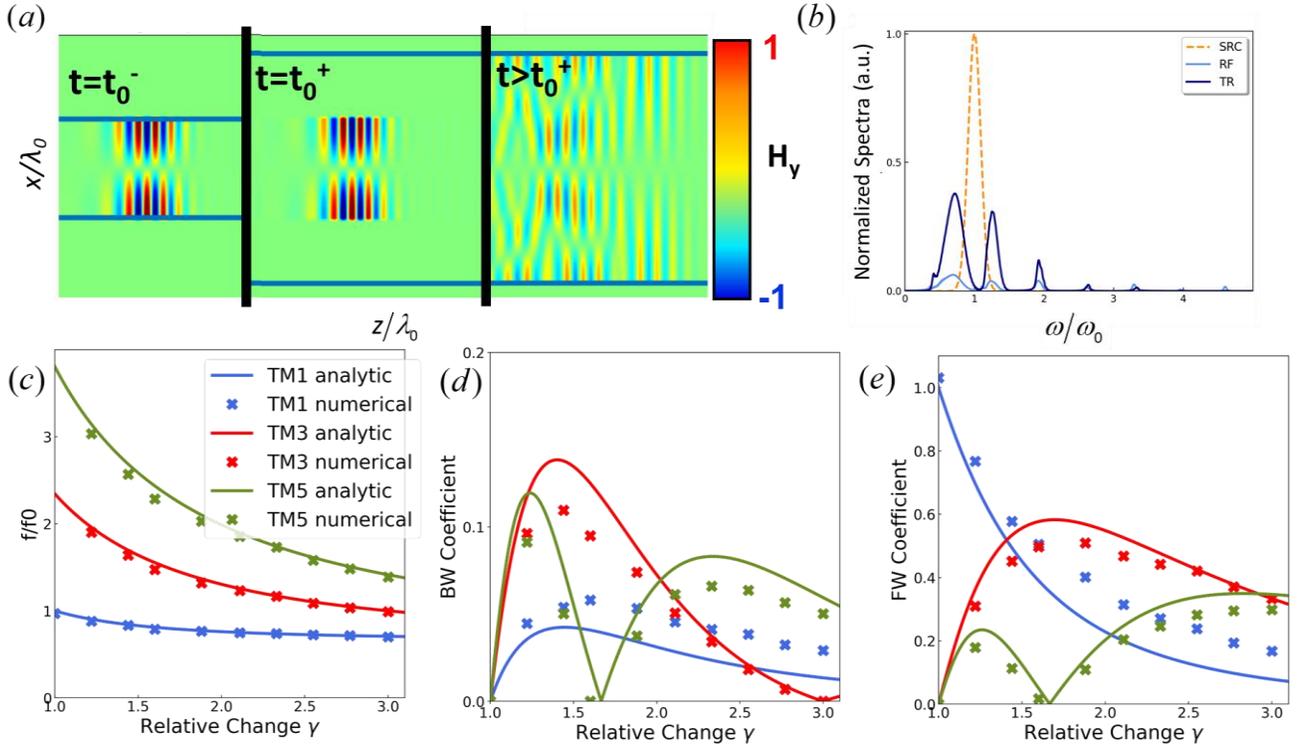

Fig. 4. (a) Snapshot in time of the $H_y$-field map within a PPWG with a boundary induced temporal interface at $t = t_0$. (b) Normalized spectra of the incident, reflected, and transmitted field at a temporal interface induced by a changing of the waveguide dimensions from $1.0d_1$ to $2d_1$. (c)-(e) Comparison between analytical and numerical results for the frequencies (c), reflection coefficients (d), and transmission coefficients (e) of the excited modes after the boundary induced temporal interface as a function of the ratio $\gamma = d_2/d_1$ between the final and the initial waveguide dimension. The legend for (c)-(e) is reported in (c).

As for the static fields, Fig. 5(a) shows the electrostatic configuration of the x- and z-components of the electric field for $t \gg t_0$. The analytical electrostatic distribution described by (7) has been compared with the numerically computed one in Fig. 5(b) for the case $\gamma = 2.5$, showing an very good agreement. Moreover, we want to verify the validity of the model to reconstruct the initial fields distribution by the superposition of the electrodynamic and electrostatic configurations, as predicted by **Error! Reference source not found.**, and (8). Fig.5(c)-(e) are devoted to this verification by reporting the analytical reconstruction of the initial spatial profile for the components $H_y, E_x, E_z$. In Fig. 5(c)-(e), the expected field profiles are indicated by using a solid red line. When the electrostatic field is not considered, the superposition of only the propagating modes after the temporal interface leads to the black dashed lines that match poorly with the expected configuration in solid red. Finally, we add the electrostatic configuration, and the blue dot-dashed line is obtained, that agrees almost perfectly with the expected profile of the filed components (solid red line). It is clear that in our analysis the presence of the electrostatic fields is a key part in describing the phenomenon.

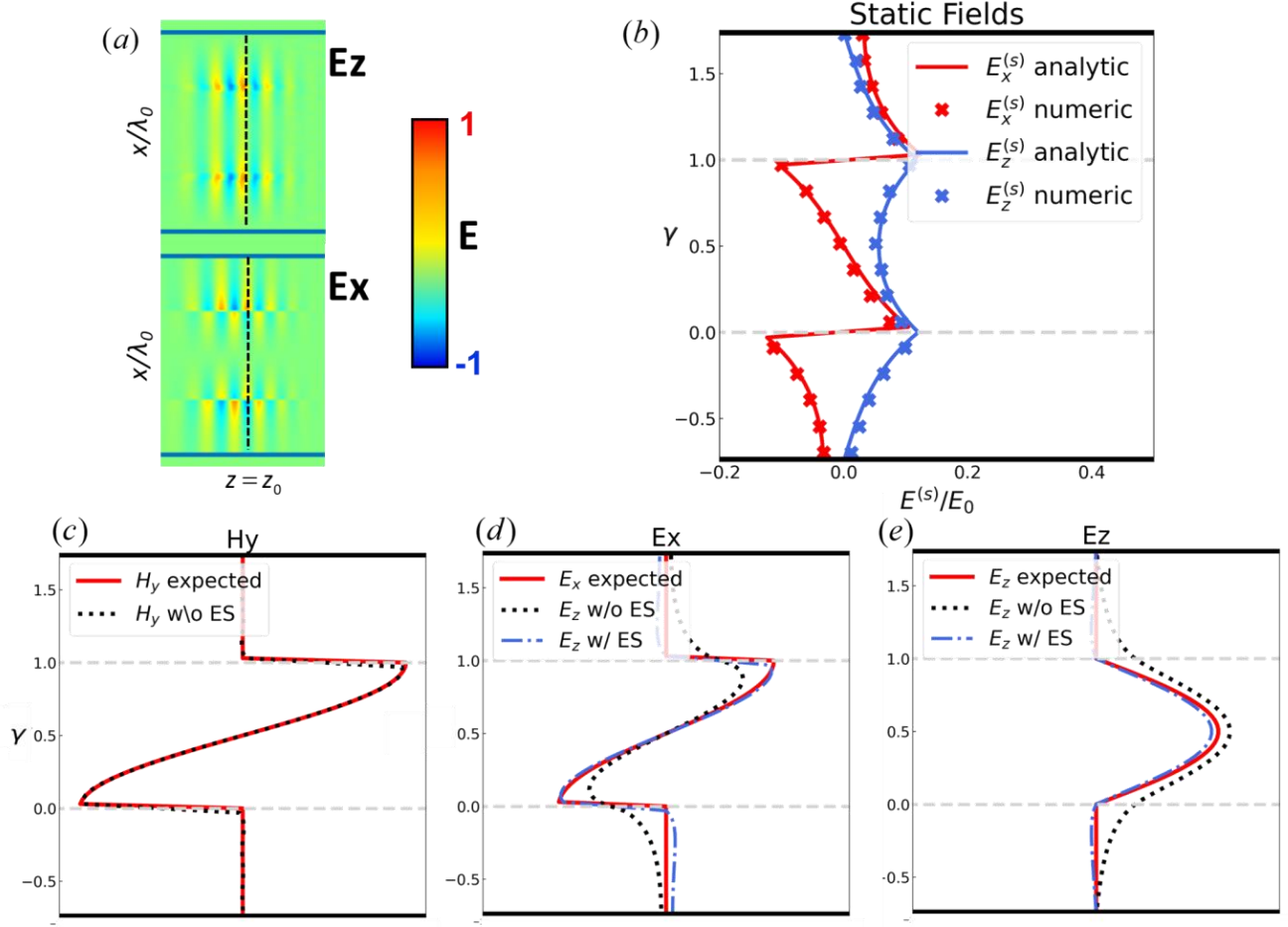

Fig.5.(a) Snapshot in time of the static fields $E^{(s)}z$ and $E^{(s)}x$. In (b) the cross-sectional view of numerical simulation of the static fields is reported, crosses, and compared with the analytical solution of the potential problem (7), solid line. In (c-d-e) the field is reconstructed without the electrostatic component (w\o ES), dotted black, and with the electrostatic (w/ ES) component, dash-dotted blue, vs the initial analytical field, solid red. It is clear that the electrostatic problem is a key part of the analysis.

# 4 CONCLUSIONS

To conclude, we have discussed the possibility to induce a temporal interface by acting on the effective medium properties rather than on the actual bulk material supporting the propagation. This has been demonstrated through the relevant example of the propagation within a parallel plate waveguide whose boundary conditions suddenly change. Our assumption led to a deeper problem regarding the nature of the electromagnetic field in its static and propagating aspect. We have described in detail the transformation from the initial to the final field configurations: we have derived analytically the frequency shifts imposed by the boundary-induced temporal interfaces and the backward and forward scattering coefficients for the propagating modes after it. Moreover, the profile of the electrostatic field raising at the interface has been derived. The analytical results have been compared to FDTD numerical results, showing a good agreement, and validating both the electrodynamic and electrostatic models. The result of this work can be generalized to other waveguide structures, *e.g.,* rectangular, dielectric waveguides, and optical fibers, or to metasurface-bounded guiding structures, enabling an easier path to design and realize novel devices working at microwave and optical frequencies based on the anomalous scattering caused by temporally switched material properties.